\documentclass[aps,prd,longbibliography,onecolumn,showpacs,amsmath,amssymb,superscriptaddress,nofootinbib,floatfix,preprintnumbers]{revtex4-1}
\usepackage[usenames,dvipsnames]{color}
\include{jdefs}
\usepackage{graphicx}
\usepackage{multirow}
\usepackage{graphicx}
\usepackage{url}
\usepackage{xspace}
\usepackage{braket}
\usepackage{tabularx}
\definecolor{darkblue}{rgb}{0,0,0.5}
\definecolor{darkgreen}{rgb}{0.1,0,0.3}
\definecolor{darkred}{rgb}{0.6,0,0}
\usepackage[bookmarks=true, bookmarksnumbered=true, colorlinks=true,
  urlcolor=darkblue, citecolor=darkgreen, linkcolor=darkred,
  breaklinks=true]{hyperref} 
\usepackage[normalem]{ulem}

\usepackage{fullpage}
\usepackage{geometry}      
\newcommand{\coherent}{CE$\nu$NS}
\begin{document}

\preprint{IPPP/18/72}
\preprint{FERMILAB-PUB-18-486-T}

\vspace*{0.7cm}

\title{How high is the neutrino floor?}
 \author{C. B\oe hm}
 \email{celine.boehm1@gmail.com }
 \affiliation{LAPTH, U. de Savoie, CNRS, BP 110, 74941 Annecy-Le-Vieux, France}
 \affiliation{School of Physics, University of Sydney, Camperdown, NSW 2006, Australia}
 \author{D.G. Cerde\~no}
 \email{d.g.cerdeno@durham.ac.uk}
 \affiliation{Institute for Particle Physics Phenomenology, Durham University, Durham DH1 3LE, United Kingdom} 
 \author{P.~A.~N. Machado}
 \email{pmachado@fnal.gov}
 \affiliation{Theory Department, Fermi National Accelerator Laboratory, P.O. Box 500, Batavia, IL 60510, USA}
 \author{A. Olivares-Del Campo}
 \email{andres.olivares@durham.ac.uk}
 \affiliation{Institute for Particle Physics Phenomenology, Durham University, Durham DH1 3LE, United Kingdom}  
 \author{E.~Perdomo}
 \email{e.perdomo-mendez@soton.ac.uk}
 \affiliation{School of Physics and Astronomy, University of Southampton, SO17 1BJ Southampton, United Kingdom}
  \author{E. Reid}
 \email{elliott.m.reid@durham.ac.uk}
 \affiliation{Institute for Particle Physics Phenomenology, Durham University, Durham DH1 3LE, United Kingdom} 
 
%  \date{\today}
 \begin{abstract}
In this paper, we compute the contribution to the coherent elastic neutrino-nucleus scattering cross section from new physics models in the neutrino sector. We use this information to calculate the maximum value of the so-called neutrino floor for direct dark matter detection experiments, which determines when these detectors are sensitive to the neutrino background. After including all relevant experimental constraints in different simplified neutrino models, we have found that the neutrino floor can increase by various orders of magnitude in the region of dark matter masses below 10~GeV in the case of scalar mediators, however, this spectacular enhancement is subject to the re-examination of supernovae bounds. The increase is approximately a factor of two for vector mediators. In the light of these results, future claims by direct detection experiments exploring the low-mass window must be carefully examined if a signal is found well above the expected Standard Model neutrino floor. 
\end{abstract}

 \maketitle
%********************

\section{Introduction}
\label{sec:introduction}

Direct detection aims at determining the nature of dark matter (DM) particles through their scattering off a target in underground detectors. A worldwide experimental effort has lead to the design and construction of extremely sensitive experiments, based on a variety of targets and techniques, which are exploring DM-nuclei interactions with unprecedented precision. A fundamental aspect in direct DM searches is the reduction of the Standard Model (SM) events in order to isolate DM ones. Modern experiments employ various techniques, such as the construction of extremely radiopure detectors, the use of a shielding, and the measurement of various channels (e.g., ionisation and scintillation) to discriminate DM signals against the background.

Given this progress, future DM experiments will soon be sensitive to a new source of background, due to coherent elastic neutrino-nucleus scattering (\coherent), which proceeds through the exchange of a $Z$ boson in the SM~\cite{Freedman:1973yd}. Neutrinos with energies in the $1-100$~MeV range are quite abundant, cannot be shielded against, and could induce keV scale nuclear recoils which would be difficult to distinguish from those caused by DM particles. For example, the recoil spectrum expected from the $^8$B solar neutrino flux would resemble that of a 6 GeV DM particle (with a specific scattering cross-section) \cite{Billard:2013qya}. This is interpreted as a ``neutrino floor''~\cite{Vergados:2008jp} in the DM-nucleus scattering cross section and DM mass parameter space, which corresponds to the threshold below which the number of neutrino events is expected to be much larger than the number of DM events, which prevents to identify DM events with certainty. Discriminating these signals would require exploiting their different contributions to annual modulation~\cite{Billard:2013qya,Davis:2014ama}, using a combination of complementary targets~\cite{Ruppin:2014bra} and  directional detectors~\cite{Grothaus:2014hja,O'Hare:2015mda} or employing detectors with improved energy resolution \cite{Dent:2016iht}.

Despite the extremely small energy deposition and weak scale cross section, \coherent\ has recently been observed by the COHERENT collaboration \cite{Akimov:2017ade} for the first time, using neutrinos from a spallation source. 
Direct detection experiments will soon be sensitive to the \coherent\ from solar neutrinos, which takes place at lower energies, thereby providing complementary information on this process. For example, the xenon based  LZ~\cite{Mount:2017qzi}, currently under construction, expects to observe approximately seven neutrino events in a three-year run.
This also offers the possibility of studying solar properties, and it is perhaps a unique opportunity to measure neutrinos from the CNO cycle \cite{Cerdeno:2017xxl,Newstead:2018muu} and thus estimate the solar metallicity \cite{Cerdeno:2017xxl}.

New physics in the neutrino sector (described in terms of new mediators between neutrinos and electrons and/or quarks or in terms of non standard effective interactions) can  increase  the \coherent\ scattering cross-section at low energies (as well as in elastic neutrino-electron scattering) \cite{Pospelov:2011ha,Harnik:2012ni,Cerdeno:2016sfi,Bertuzzo:2017tuf,Dutta:2017nht,AristizabalSierra:2017joc,Coloma:2017egw,Gonzalez-Garcia:2018dep,AristizabalSierra:2018eqm,Kosmas:2017tsq,Papoulias:2018uzy,Billard:2018jnl,Denton:2018xmq}. This can raise the neutrino floor, inevitably affecting the search for light DM particles in upcoming experiments, especially in those that will explore the low-mass DM window.  In this article we study a range of simplified models with emphasis on low-mass mediators in the neutrino sector to which we apply the most recent constraints in order to determine how high the neutrino floor can be. This information is crucial in order to correctly interpret a future signal in direct DM detectors.

We have found that  the \coherent\ floor can be raised by several orders of magnitude in the region with DM mass below 10~GeV when a new scalar mediator is assumed. However, the impact that such a mediator could have on the equation of state of a supernovae core would require further analysis. The increase is only by a factor of approximately $ 2$ for a new vector mediator. This result already affects the background predictions for xenon based experiments such as XENON1T \cite{Aprile:2018dbl} and, more importantly, it has deep implications for future results from detectors that explore the low-mass DM window, such as SuperCDMS SNOLAB \cite{Agnese:2016cpb} and NEWS-G \cite{Arnaud:2017bjh}.

This article is organised as follows. In Section~\ref{sec:floor}, we review the computation of the neutrino floor in the SM and introduce various simplified models for new physics in the neutrino sector. For each of them, we review the main experimental constraints and, using these, we evaluate the highest possible neutrino rate that can be expected in direct DM detection experiments. In Section~\ref{sec:results} we obtain a new (\emph{raised}) neutrino floor that we put into the context of the next generation dark matter detectors. In Section~\ref{sec:conclusions} we present the conclusions of this work.

\section{New physics in the neutrino sector and the \coherent\ floor}
\label{sec:floor}

The total number of expected events from \coherent\ in a direct DM detection experiment with a given exposure, $\epsilon$, can be computed by integrating the  \coherent\ cross section, $d \sigma_{\nu N}/dE_R$, and the incoming flux of neutrinos, $d\phi_{\nu }/dE_\nu$, over the resulting nuclear recoil energy, 
\begin{equation}
N_{{\rm CE\nu NS}} = \frac{\epsilon}{m_N} \int_{E_{T}}^{E_{\mathrm{max}}} dE_R \int_{E_{\nu}^{\mathrm{min}}} dE_\nu \frac{d\phi_\nu}{dE_\nu} \frac{d\sigma_{\nu N}}{dE_R},
\label{eq:Neutrino_rate}
\end{equation}
where $m_N$ is the nuclear mass, and $E_\nu^{\mathrm{min}}$ is the minimum neutrino energy to produce a nuclear recoil of energy $E_R$.
In the SM, the coherent scattering of neutrinos off nuclei takes place through the exchange of a $Z$ boson, and the resulting cross section reads \cite{Freedman:1973yd}
\begin{equation}
\frac{d\sigma_{\nu N}}{dE_R} = \frac{G_F^2}{4 \pi} Q_{vN}^{2} m_N \left(1-\frac{m_NE_R}{2E_\nu^2}\right) F^2(E_R) ,
\label{eq:nuesigma}
\end{equation}
where $G_F$ is the Fermi constant, $Q_{vN}=N-(1-4\sin^2 \theta_w)Z$ is the weak hypercharge of a target nucleus containing $N$ neutrons and $Z$ protons, and $F^2(E_R)$ is the nuclear form factor, for which we have taken the parametrisation given by Helm \cite{Helm:1956zz}. The scattering cross section benefits from a coherence factor that scales as the total number of nucleons squared $Q_{vN}^{2}\sim A^2$. The neutrino flux at low energies is dominated by solar neutrinos, and the relevant fluxes used in this work can be found in Refs.~\cite{Bahcall:2004mq,Vinyoles:2017bqj}. At higher energies, atmospheric neutrinos are the most important source, although their flux is substantially smaller \cite{Battistoni:2005pd}. 
As a side note, although there are systematic and statistical uncertainties of the order of $1-10\%$ associated to the flux of solar neutrinos, we will neglect these in calculating the neutrino floor. As we will show below, the effect of new physics can be much larger than such uncertainties.

We construct the neutrino floor as follows, based on Ref.~\cite{Billard:2013qya}. For a given target nucleus, a minimum energy threshold $E_T$ is set and, using Eq.~(\ref{eq:Neutrino_rate}), the exposure required to give 1 expected count of \coherent\ is calculated. Using this value of the exposure, one can compute the minimum spin-independent WIMP-nucleon elastic cross section, $\sigma^{SI}_{\chi n}$, that can be excluded at the 90\% confidence level for each value of the DM particle mass, $m_\chi$. For a background-free analysis, this lies along the 2.3 DM event isovalue contour. The threshold energy is then varied across the relevant range and, by taking a lower envelope on $\sigma^{SI}_{\chi n}$, we obtain the contour in parameter space along which, given an optimal choice of the threshold to minimise the neutrino background, there will be as many \coherent\ events as WIMP events. Alternatively, it is also possible to define the neutrino floor as a DM discovery limit using spectral information and including uncertainties in the solar neutrino fluxes \cite{Billard:2013qya}. The neutrino floor can also be generalised to other types of DM-nucleus effective field theory operators \cite{Dent:2016wor}, but in this work we assume only a spin-independent WIMP-nucleon interaction. In our calculation of the DM signature we have assumed a Standard Halo Model with a local density of 0.4~GeV~cm$^{-3}$, a central velocity of $230$~km~s$^{-1}$, and a velocity dispersion of $156$~km~s$^{-1}$.

As mentioned in the Introduction, disentangling DM and neutrino signals in the region of parameter space below this line is not impossible, but the \coherent\ floor serves as an indication of the point at which neutrinos become a significant obstacle to DM direct detection. New physics in the neutrino sector can contribute to the predicted \coherent\ cross section, thus shifting the neutrino floor. These contributions are larger for light mediators \cite{Cerdeno:2016sfi,Dent:2016wor,Lindner:2016wff,AristizabalSierra:2017joc,Bertuzzo:2017tuf,Papoulias:2018uzy}.

\subsection{Models}
\label{sec:models}

We have considered a set of low scale simplified models in which the SM structure is extended by the inclusion of a new light mediator \cite{Boehm:2002yz,Boehm:2004uq} between the neutrinos and quarks (and/or leptons). An obvious concern of dealing with simplified models at low scales is the difficulty in realizing such models in UV complete frameworks. This concern is justified as such models typically have chiral anomalies, requiring extra light fermion content to fix it, or non-trivial scalar sectors associated to the breaking of some symmetry at low scales. To mitigate this worry, we will focus on low scale simplified models that may have a clear UV completion: gauged $B-L$~\cite{Nelson:2007yq, Harnik:2012ni}, gauged $B-L(3)$ of the third family~\cite{Babu:2017olk}, sequential $Z'$~\cite{Davoudiasl:2014kua} and scalar mediators (see e.g. Ref.~\cite{Gabriel:2006ns}).

\begin{itemize}
\item {\bf Vector/Axial Vector Mediator:}

The introduction of a new vector field, $Z'$, that  couples to SM fermions gives rise to new terms in the SM Lagrangian of the form
\begin{equation}
\mathcal{L} \supset - \left(g_{Z'} J^\mu_{Z'} - \frac{g}{c_W}\epsilon'J^\mu_{\rm Z}- e\epsilon J^\mu_{\rm em}\right)Z'_\mu \ ,
\label{eq:Vector_Lagrangian}
\end{equation}
where $g_{Z'}$ is the gauge coupling of the new gauge group; $J_{Z'}$, $J_{\rm em}$, and $J_{Z}$ are the $Z'$, electromagnetic, and $Z$ currents; and $\epsilon$ and $\epsilon'$ parametrize the $Z'$ mixing with the photon and the $Z$ boson, respectively. Here we will not study any model with kinetic mixing\footnote{If the $U(1)$ studied here is a subgroup of a non-Abelian gauge group, kinetic mixing is forbidden at tree level, although it will be induced at loop level. The loop contribution depends on the fermion content of the UV theory, but generically we expect it to be suppressed by a factor of $\epsilon\sim g_{B-L}e/16\pi^2\sim2\times 10^{-3} g_{B-L}$, and is therefore negligible.}, so we can disregard the last term in eq.~(\ref{eq:Vector_Lagrangian}). To ease the notation, we parametrise the Lagrangian as
\begin{equation}
\mathcal{L} \supset - \sum_{f} c_{f} \bar f \gamma^\mu f Z'_\mu+{\rm h.c.} \ ,
\label{eq:Vector_Lagrangian_simplified}
\end{equation}
where the sum runs over all left- and right-handed fermion fields, that is $f=Q_L,u_R,d_R,L,e_R$ for each flavour. For the $B-L$ case, $c_{f}=g_{B-L}/3$ for quarks and $c_{f}=-g_{B-L}$ for leptons. In the sequential $Z'$, all couplings come from the mass mixing to the SM $Z$ boson, $\epsilon'$, and thus $c_{f}$ are given by $g_{Z'}\epsilon'$ times the  $Z$ couplings of each fermion. In the $B-L(3)$ model, the couplings to the third family are identical to the $B-L$, while the coupling to the first two comes from $Z-Z'$ mass mixing.
The resulting \coherent\ cross section can be written as 
\begin{equation}
\frac{d\sigma_{\nu N}}{dE_R} = \frac{d\sigma^{SM}_{\nu N}}{dE_R} - \left(\frac{G_F m_N Q_{\nu N}  Q_{\nu N, v}' (2E^2_{\nu}-E_R m_N)}{2\ \sqrt[]{2} \pi E^2_{\nu} \left(2E_R m_N+m^2_{Z'}\right)} - \frac{Q_{\nu N, v}'^2 m_N (2E^2_{\nu}-E_R m_N)}{4 \pi E^2_{\nu} \left(2E_R m_N+m^2_{Z'}\right)^2} \right)F^2(E_R) ,
\label{eq:Vector_sigma}
\end{equation}
%\begin{equation}
%\frac{d\sigma_{\nu N}}{dE_R} = \frac{d\sigma^{SM}_{\nu N}}{dE_R} - \frac{G_F m_N Q_{\nu N}  Q_{\nu N, v}' (2E^2_{\nu}-E_R m_N)}{2\ \sqrt[]{2} \pi E^2_{\nu} \left(2E_R m_N+m^2_{Z'}\right)} F^2(E_R)  + \frac{Q_{\nu N, v}'^2 m_N (2E^2_{\nu}-E_R m_N)}{4 \pi E^2_{\nu} \left(2E_R m_N+m^2_{Z'}\right)^2} F^2(E_R) ,
%\label{eq:Vector_sigma}
%\end{equation}
where the SM cross section is given in Eq.~(\ref{eq:nuesigma}).  Here $Q_{\nu N}$ and $Q_{\nu N, v}'$ are the coherence factors of the cross section, the latter being given by
\begin{eqnarray}
Q_{\nu N, v}' &=& \left[(2Z+N)\frac{(c_{Q_L}+c_{u_R})}{2} + (Z+2N)\frac{(c_{Q_L}+c_{d_R})}{2}\right] c_{\nu}\ .
\label{eq:VectorQ'}
\end{eqnarray}
Eq.(\ref{eq:Vector_Lagrangian_simplified})  assumes a vector mediator. However we did check the case of an axial coupling. Typically, axial interactions contribute less significantly to the \coherent~cross section than vector interactions, as the former couple to the overall spin of the nucleus \cite{Alarcon:2011zs,Alarcon:2012nr,DelNobile:2013sia,Hill:2014yxa}.
The coherence factor for an axial interaction is proportional to the nuclear angular momentum,
and does not benefit from the $\sim A^2$ enhancement. Since the couplings $c_\nu$ are still affected by the constraints from electron interactions, one should not expect a large contribution from the axial component for heavy nuclei. However, this contribution can be significant for light targets provided they have non-vanishing nuclear angular momentum. In our study, we have considered Ge and Xe (which are heavy targets), and He (which has zero spin), for all of which the contribution from axial couplings is negligible, and thus has been dropped out in Eq.~(\ref{eq:VectorQ'}).

To obtain the \coherent~cross section for any of the models considered here, we simply need to identify the corresponding $c_f$ couplings. Different models have different couplings to quarks and leptons, leading to distinct constraints on the values of the gauge coupling and mediator mass: the constraints used in this paper for the $B-L(3)$ model are taken from Ref.~\cite{Babu:2017olk,Ilten:2018crw,Bauer:2018onh} for the case $\tan\beta=10$, which leads to $\epsilon'\simeq0.01g_{B-L(3)}$; while the constraints on the $B-L$ model are a combination of those used in Refs.~\cite{Harnik:2012ni, Bilmis:2015lja, Rrapaj:2015wgs,Cerdeno:2016sfi} and Big Bang nucleosynthesis (BBN) constraints given in Ref \cite{Huang:2017egl}. The Sequential SM turns out to be extremely constrained and the resulting contribution to the neutrino floor is very small, thus we will not discuss it further.

\item{\bf Scalar/Pseudoscalar mediator:}

The other scenario of interest which may impact the neutrino floor is constituted by a  light scalar mediator that interacts with SM fermions \cite{Boehm:2003hm,Boehm:2006mi}. We consider here a simple extension of the form
\begin{equation}
\mathcal{L} = -y_\nu  \bar{\nu}_L^c \phi \nu_L - \sum_{f\neq\nu}y_f\bar{f} \phi f  - \sum_{f\neq\nu}y_f^5\bar{f} \phi i\gamma_5 f  + {\rm h.c.}\ ,
\label{eq:Scalar_Lagrangian}
\end{equation}
where the sum runs over all charged fermions. Note that in this scenario the scalar coupling violates lepton number\footnote{One could also work with a lepton number conserving model, at the expense of including right-handed neutrinos. The predictions for \coherent\ would not change but this scenario is more affected by supernova constraints, which limit the contribution to the neutrino floor}. For simplicity, we assume that all SM particles have the same coupling $y_f=y$ to $\phi$. We also neglect $y_f^5$, as this pseudoscalar coupling leads to a very small contribution to the coherent scattering cross section (see e.g.~\cite{Hill:2014yxa}). The resulting CE$\nu$NS cross section reads
\begin{equation}
\frac{d\sigma_{\nu N}}{dE_R} = \frac{d\sigma^{SM}_{\nu N}}{dE_R} + \frac{y^4 Q_{\nu N, s}'^2 m_N^2 E_R}{4 \pi E_\nu ^2 (2 E_R m_N + m_\phi ^2)^2}F^2(E_R) \ ,
\label{eq:Scalar_sigma}
\end{equation}
where $m_\phi$ is the mass of the scalar mediator and the new coherence factor $Q_{\nu N, s}'=13.8A-0.02Z$ is computed using Refs.~\cite{Crivellin:2013ipa,Junnarkar:2013ac,Hoferichter:2015dsa,Ellis:2018dmb} to calculate the scalar-quark form factors.

Compared with the models with a vector mediator discussed above, the specific couplings of this scalar model are less motivated by theory. It therefore has fewer model specific constraints. In this work, we have considered the bounds from astrophysical and cosmological sources discussed in Ref.~\cite{Farzan:2018gtr}, and the results of the COHERENT experiment \cite{Akimov:2017ade}.

\end{itemize}

\begin{figure}[t!] 
\includegraphics[width=7.8cm]{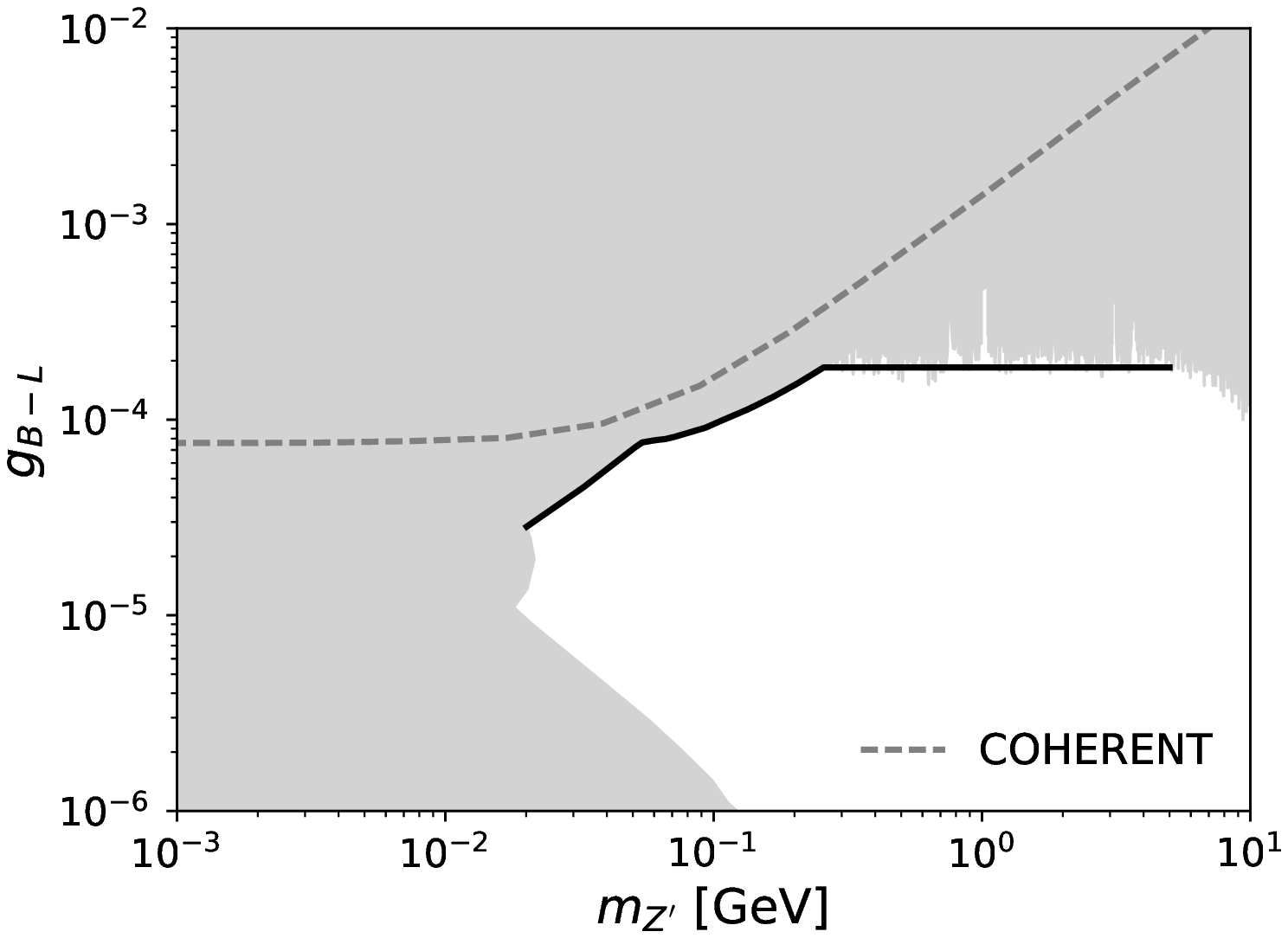}
\includegraphics[width=7.8cm]{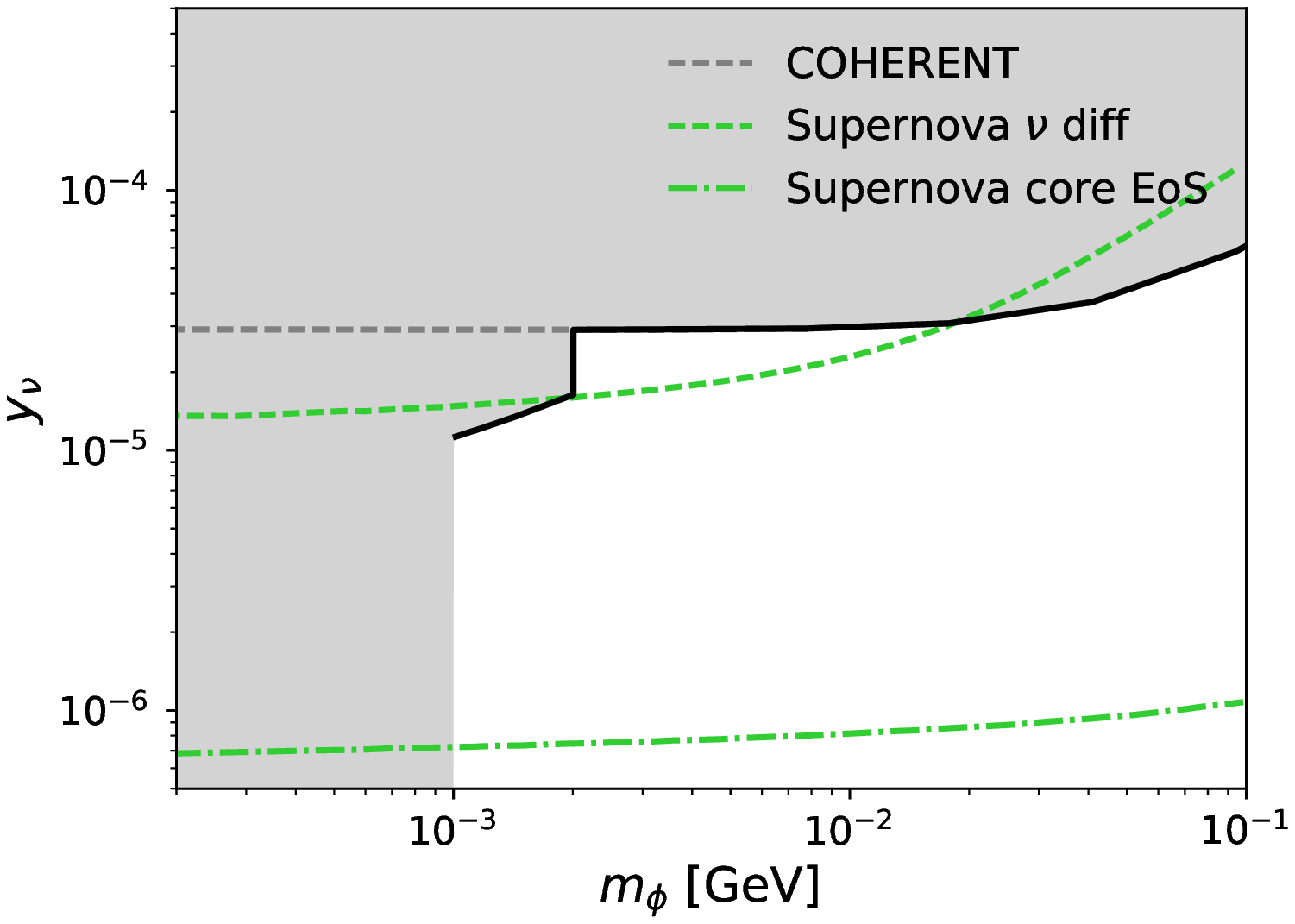}
\caption{The gray areas represent the regions excluded in the case of a vector (left) and scalar (right) mediator. The solid black line represents the values of the mediator coupling that we have used to determine the maximum contribution to the neutrino floor in both cases. In the scalar case we include constraints from supernovae in green: both those from neutrino diffusion (dashed) and the core equation of state (dot-dashed). }
\label{fig:constraints}
\end{figure}

Fig.~\ref{fig:constraints} represents the areas in the mediator mass and coupling parameter space that are available for models with new vector (left) and scalar (right) mediators. Gray regions are excluded by various astrophysical and cosmological limits, as well as bounds from neutrino experiments. The upper bound on new physics couplings obtained from the COHERENT observation \cite{Akimov:2017ade} is shown by means of a dashed gray line \cite{Liao:2017uzy,Kosmas:2017tsq,Farzan:2018gtr}. The green dashed and green dot-dashed lines for scalar mediators indicate the values of the neutrino coupling for which the neutrino diffusion rate and the core equation of state in supernovae could be significantly altered and need to be reevaluated. 
Vector mediators are extremely constrained by a combination of bounds from neutrino experiments (mainly Borexino, GEMMA) as well as astrophysical constraints (on supernovae and other stellar systems).   
Contrariwise, models with extra scalar mediators are in principle more flexible (with the caveat that supernovae limits might have to be reevaluated).

\section{Results}
\label{sec:results}

For each simplified model described in Sec.~\ref{sec:floor}, we have considered the largest possible values of the neutrino couplings as a function of the mediator mass that is allowed by the various experimental constraints (represented as a solid black line in Fig.~\ref{fig:constraints}), and we have used these to determine the maximum contribution to the \coherent\ cross section. The height and shape of the \coherent\ floor vary for different target nuclei. Here we consider three different materials. Germanium and xenon have qualitatively similar shapes, but we include both as they are common targets in low and high mass searches respectively, including SuperCDMS~\cite{Agnese:2016cpb}, XENON1T~\cite{Aprile:2018dbl}, LZ~\cite{Mount:2017qzi}, and DARWIN \cite{Aalbers:2016jon}. We also include helium, as an example of a very light target, which has been proposed as a way of probing very low DM masses in a future phase of the NEWS-G experiment~\cite{Arnaud:2017bjh}. The sensitivity line for NEWS-G has been extracted from Ref.~\cite{Katsioulas:2018}. The very low mass of the He nucleus allows solar $^8$B neutrinos to generate much higher energy recoils. The resulting flattening of the recoil spectrum prevents us from distinguishing $^8$B neutrinos from higher mass DM simply by choosing a higher energy threshold, and so the neutrino floor is noticeably flatter than it is for heavier targets.

\begin{figure}[t!]
\includegraphics[width=16cm]{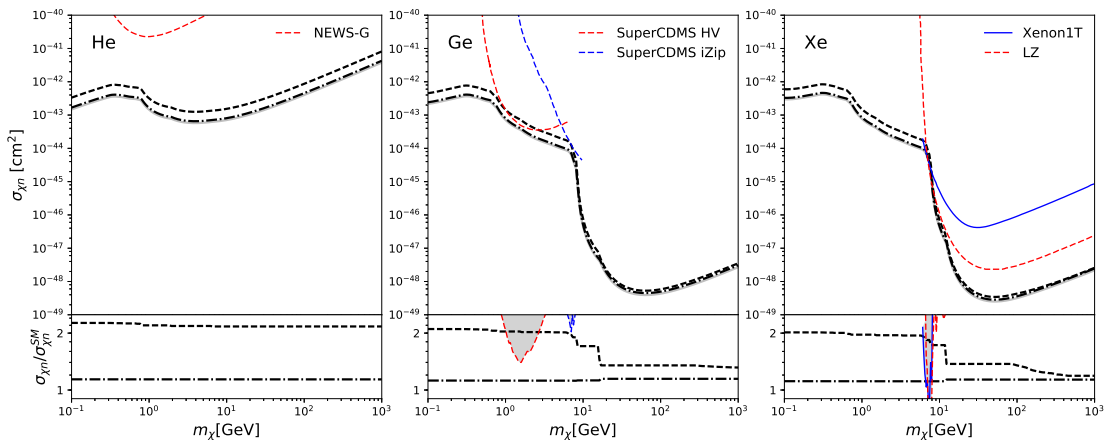}
\caption{
Upper: \coherent\ floor for a new vector mediator, computed for direct detection experiments utilising, from left to right, He, Ge, and Xe. The SM neutrino floor (solid, grey) is compared with the maximum level reached in a $B-L$ (dashed, black) and a $B-L(3)$ (dot-dashed, black) model. For comparison, the sensitivities of some current (solid) and future (dashed) direct detection experiments are shown in color.  Lower: Ratio of the new neutrino floor to the SM result. The sensitivities of representative direct detection experiments are also shown in this parameter space.
}
\label{fig:vector_floor}
\end{figure}

Figure \ref{fig:vector_floor} represents the resulting \coherent\ floor for the two vector mediated models discussed in Sec \ref{sec:models}. For comparison, the SM contribution is shown as a solid grey line. We can observe that the new physics contribution can be greater than a factor of 2 for DM masses below 10~GeV. The $B-L$ model (black dashed line) has a greater enhancement at low masses than the $B-L(3)$ (black dot-dashed line) due to less stringent constraints on the mediator mass. However, at higher energies the $B-L(3)$ enhancement is comparable, as larger couplings to the third generation are allowed with higher mediator masses. We also observe that current direct detection experiments are beginning to probe the region of parameter space below the ``new'' neutrino floor, suggesting that future detectors could be used to put competitive limits on the properties of these new vector mediators. It should be noted that astrophysical uncertainties in the parameters that describe the DM halo can lead to a greater effect in the DM discovery limit over the neutrino floor \cite{OHare:2016pjy}.

\begin{figure}[t!]
\includegraphics[width=16cm]{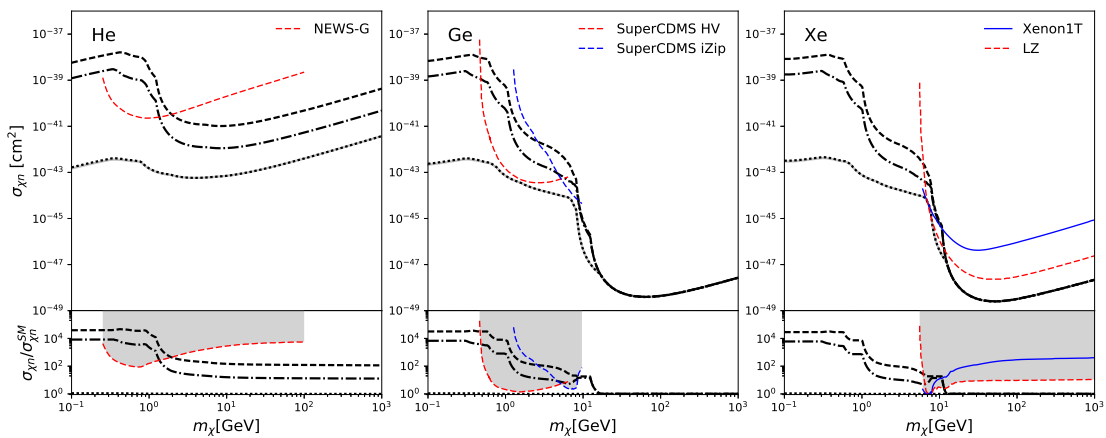}
\caption{
As in Fig. \ref{fig:vector_floor}, but for a scalar mediator. For the constraints on our model, we consider three different cases discussed in Sec~\ref{sec:models}: one in which supernova constraints are neglected (dashed), one in which supernova diffraction constraints are included but bounds from the SN core EoS are ignored (dot-dashed) and one in which all supernova constraints are included (dotted).
}
\label{fig:scalar_floor}
\end{figure}

As expected, models with scalar mediators allow for a much larger enhancement of the neutrino floor, represented by a dashed line in Fig.~\ref{fig:scalar_floor}. However, the spectacular increase of several orders of magnitude for DM masses below 10~GeV is subject to the reevaluation of supernovae constraints in this kind of lepton-violating models. As pointed out in Ref.~\cite{Farzan:2018gtr}, it is uncertain whether this range of mediator masses and couplings can induce changes in the equation of state that describes the supernova core and the physics of neutrino diffusion. To account for these effects, in Fig.~\ref{fig:scalar_floor} we also show the results when neutrino diffusion are limits included (dot-dashed line) and when a strict limit on  the supernova core equation of state is also added (dotted line). The spectacular enhancement of the neutrino floor at small DM masses corresponds to very light new mediators (with masses in the MeV range) \cite{Boehm:2003hm,Boehm:2006mi}, while for heavier mediators, such as those considered in Ref.~\cite{Bertuzzo:2017tuf}, the increase is much more moderate.

The new scalar mediator gives very little enhancement to the neutrino floor at higher WIMP masses, since the region of heavy mediators is more constrained from particle physics bounds, meaning that the best prospects to constrain such models come from experiments with low energy thresholds such as SuperCDMS SNOLAB.

The shape and height of the neutrino floor depend on the nature of the DM interaction, and thus they change significantly for different EFT operators \cite{Dent:2016wor}, especially when these feature non-trivial momentum or velocity dependence. However, we have checked explicitly that the ratio by which the neutrino floor is raised with respect to its SM value is only slightly distorted. In particular, the maximum increase in the neutrino floor at small DM masses is insensitive to the choice of EFT operator. Therefore, the lower panels of Figures~\ref{fig:vector_floor} and \ref{fig:scalar_floor} are a useful guide to the results for all other EFT operators. 

\section{Conclusions}
\label{sec:conclusions}

In this paper we have determined the contribution from new physics models to the coherent neutrino scattering (\coherent) floor, which is expected to be within the reach of next-generation DM direct detection experiments. We have considered a collection of simplified models that include a new vector or scalar mediator between the SM neutrino and the SM quarks and leptons. We have incorporated the most recent constraints from various sources of experiments and astrophysical observations and used them to determine the maximum reach of the neutrino floor in the parameter space of elastic spin-independent DM scattering. In doing this, we have payed particular attention to the limits on new physics that can be derived from the recent observation of \coherent\ by the COHERENT collaboration.

We have observed that, in the case of vector mediators embedded in UV complete frameworks, the \coherent\ floor can be raised by approximately a factor of two for small DM masses (below 10~GeV, where the main contribution is due to solar neutrinos) and by a factor of 1.3 for large DM masses (where atmospheric neutrinos dominate). Experimental limits from neutrino and beam dump experiments are the main obstacle that limits the height of the neutrino floor in these scenarios.

In the case of new scalar mediators, the neutrino floor can be raised by several orders of magnitude in the region of low-mass DM (below 10~GeV), a feature that is definitely within the reach of upcoming experiments such as SuperCDMS SNOLAB and NEWS-G. However, this spectacular enhancement is subject to the re-examination of supernovae bounds, as new physics can induce changes in the equation of state of the supernova core that must be carefully analysed. If these bounds turn out to be as strong as suggested in Ref.~\cite{Farzan:2018gtr}, the maximum enhancement of the neutrino floor due to a light scalar mediator would be quite small.

In conclusion, our results indicate that the expected \coherent\ background in the recent XENON1T results could increase by an a factor of two or even more. 
More importantly, future claims by DM experiments in the low-mass window must be carefully examined to discriminate neutrino and DM signals well above the expected SM neutrino floor.

\acknowledgements

We are grateful to Bhaskar Dutta, Yasaman Farzan, Kostas Nikolopoulos, and Louis Strigari for useful discussions. DC and ER thank Fermilab, and PM and AO thank IFT-Madrid, for kind hospitality during the completion of this project. We acknowledge support from the Marie Sk\l{}odowska-Curie grant agreement No. 690575 (RISE InvisiblesPlus). AO is supported by the  European  Research  Council  under  ERC  Grant ``NuMass''  (FP7-IDEAS-ERC  ERC-CG  617143). Fermilab is operated by the Fermi
Research Alliance, LLC under contract No. DE-AC02-07CH11359 with the United States Department of Energy.

%********************

\providecommand{\href}[2]{#2}\begingroup\raggedright\endgroup

\end{document}